\documentclass[11pt]{article}

\usepackage[margin=1in]{geometry}
\usepackage{amsmath, amssymb, amsfonts}
\usepackage{graphicx}
\usepackage{booktabs}
\usepackage{bm}
\usepackage{setspace}
\usepackage{color, xcolor}
\usepackage{makecell}
\usepackage{natbib}
\usepackage{caption}
\usepackage{subcaption}
\usepackage{authblk}
\usepackage{alltt}

\usepackage[ruled, noend]{algorithm2e}

\usepackage{csquotes}
\MakeOuterQuote{"}

\usepackage{hyperref}
\hypersetup{
    colorlinks,
    linkcolor={red!50!black},
    citecolor={blue!50!black},
    urlcolor={blue!80!black}
}

\makeatletter
\def\maxwidth{ %
  \ifdim\Gin@nat@width>\linewidth
    \linewidth
  \else
    \Gin@nat@width
  \fi
}
\makeatother

\usepackage{framed}
\makeatletter
\newenvironment{kframe}{%
 \def\at@end@of@kframe{}%
 \ifinner\ifhmode%
  \def\at@end@of@kframe{\end{minipage}}%
  \begin{minipage}{\columnwidth}%
 \fi\fi%
 \def\FrameCommand##1{\hskip\@totalleftmargin \hskip-\fboxsep
 \colorbox{shadecolor}{##1}\hskip-\fboxsep
     \hskip-\linewidth \hskip-\@totalleftmargin \hskip\columnwidth}%
 \MakeFramed {\advance\hsize-\width
   \@totalleftmargin\z@ \linewidth\hsize
   \@setminipage}}%
 {\par\unskip\endMakeFramed%
 \at@end@of@kframe}
\makeatother

\definecolor{fgcolor}{rgb}{0.345, 0.345, 0.345}
\newcommand{\hlnum}[1]{\textcolor[rgb]{0.686,0.059,0.569}{#1}}%
\newcommand{\hlstr}[1]{\textcolor[rgb]{0.192,0.494,0.8}{#1}}%
\newcommand{\hlcom}[1]{\textcolor[rgb]{0.678,0.584,0.686}{\textit{#1}}}%
\newcommand{\hlopt}[1]{\textcolor[rgb]{0,0,0}{#1}}%
\newcommand{\hlstd}[1]{\textcolor[rgb]{0.345,0.345,0.345}{#1}}%
\newcommand{\hlkwb}[1]{\textcolor[rgb]{0.69,0.353,0.396}{#1}}%
\newcommand{\hlkwc}[1]{\textcolor[rgb]{0.333,0.667,0.333}{#1}}%
\newcommand{\hlkwd}[1]{\textcolor[rgb]{0.737,0.353,0.396}{\textbf{#1}}}%

\newenvironment{knitrout}{}{}

\title{Flexible yet Sparse Bayesian Survival Models with Time-Varying Coefficients and Unobserved Heterogeneity}
\author[1,2]{Peter Knaus}
\author[3]{Daniel Winkler}
\author[4]{Sebastian F. Schoppmann}
\author[4]{Gerd Jomrich}
\affil[1]{Department of Statistics, Harvard University}
\affil[2]{Institute for Statistics and Mathematics, Vienna University of Economics and Business}
\affil[3]{Institute for Retailing \& Data Science, Vienna University of Economics and Business}
\affil[4]{Department of Surgery, Comprehensive Cancer Center Vienna,  Medical University of Vienna}

\date{}

\begin{document}

\maketitle

\begin{abstract}
Survival analysis is an important area of medical research, yet existing models often struggle to balance simplicity with flexibility. Simple models require minimal adjustments but come with strong assumptions, while more flexible models require significant input and tuning from researchers. We present a survival model using a Bayesian hierarchical shrinkage method that automatically determines whether each covariate should be treated as static, time-varying, or excluded altogether. This approach strikes a balance between simplicity and flexibility, minimizes the need for tuning, and naturally quantifies uncertainty. The method is supported by an efficient Markov chain Monte Carlo sampler, implemented in the R package shrinkDSM. 
Comprehensive simulation studies and an application to a clinical dataset involving patients with adenocarcinoma of the gastroesophageal junction showcase the advantages of our approach compared to existing models.
\end{abstract}

\noindent\textbf{Keywords:} State-space model, Hierarchical shrinkage priors, Piecewise-exponential model, R package

\section{Introduction}\label{sec:intro}

Survival analysis often deals with data on long timescales. Medical data provides a natural example: patients are often observed for several years or even decades during a study. In such cases, assuming that covariate effects on the hazard rate remain constant over time may lead to model misspecification. Conversely, allowing all coefficients to vary freely over time can lead to overfitting, introducing spurious variation that obscures the true covariate effects on survival. Thus, models should be able to automatically determine not only which covariates exhibit time-varying effects but also which should remain constant or be excluded entirely, striking a balance between flexibility and parsimony.

The two most widely used approaches to survival analysis fall on opposite ends of this flexibility spectrum. On one end, the Cox proportional-hazards model \citep{cox1972regression} assumes that covariate effects on the hazard rate are constant over time, providing a parsimonious approach for modeling survival. At the other extreme, the Kaplan-Meier estimator \citep{kaplan1958nonparametric} provides complete flexibility by estimating survival probabilities at each event time without parametric assumptions. However, it does not model covariate effects directly, particularly in the case of  continuous or multiple covariates, and can struggle with stratification in high-dimensional settings.

Existing work on balancing flexibility and parsimony in survival models has followed two main approaches: one focusing on variable selection and the other on modeling time variation. The first approach focuses on selecting relevant covariates without modeling time variation, using either penalized likelihood methods \cite{tibshirani1997lasso, fan2002variable, wu2012elastic} or Bayesian variable selection techniques \cite{faraggi1998bayesian, sha2006bayesian, newcombe2017weibull}. The second approach increases model flexibility by incorporating time-varying effects. Prominent examples include the additive model of \cite{aalen1980model}, piecewise exponential hazard models \cite{gamerman1991dynamic, wagner2011bayesian}, and various spline-based approaches \cite{gray1992flexible, perperoglou2014cox}. Attempts have also been made to unify these approaches, allowing for both variable selection and time-varying effects \citep{yan2012model}. However, these flexible methods often require extensive hyperparameter tuning and may even necessitate pre-specifying which variables vary over time \cite{perperoglou2014cox}, a decision researchers may not be able to make a priori.

We propose a Bayesian hierarchical shrinkage-based method that draws on methods developed for time-varying parameter models to automatically determine whether covariates should be static, time-varying, or excluded entirely. Our approach offers several advantages. First, it performs both variable selection (identifying relevant covariates) and variance selection (determining which effects are time-varying), effectively striking a balance between flexibility and model simplicity. Second, unlike many existing methods, it requires minimal hyperparameter tuning, as all key parameters are inferred from the data. The only exception is the grid size for the time points at which the parameters can vary, which can be selected in a straightforward, data-driven manner. Parameters can only vary at observed exit points, and the Bayesian hierarchical shrinkage approach is effective even when parameters vary with each observed exit. (see Section~\ref{sec:model} for details). Third, our fully Bayesian formulation provides a natural framework for quantifying uncertainty, enhancing both predictions and interpretability.

To facilitate estimation, we develop an efficient Markov chain Monte Carlo (MCMC) sampler and implement our method in the \texttt{R} package \texttt{shrinkDSM}, providing a user-friendly tool for researchers seeking a flexible time-to-event model with minimal tuning. Additionally, \texttt{shrinkDSM} enables sampling from the posterior predictive distribution of survival times given new covariates. This is a key advantage of absolute risk models over relative risk models, as it allows for direct estimation of survival probabilities rather than hazard ratios. The package also supports time-varying covariates as inputs and allows for the specification of a time-varying grouped factor model (see \autoref{sec:factor}), further extending its applicability. While the package provides helper functions for these features in R, the MCMC is entirely implemented in C++, ensuring fast and efficient sampling.

To illustrate the broad applicability of our proposed model, we benchmark it against 8 other models using a comprehensive simulation study. Additionally, we apply the model to a recent dataset involving patients with adenocarcinoma of the gastroesophageal junction, specifically analyzing those who underwent either primary resection or neoadjuvant therapy prior to surgery. This application highlights how our model enhances the understanding of the effects of covariates on survival.

The remainder of the paper is structured as follows: Section~\ref{sec:model} describes the model components and outlines the MCMC estimation scheme. Section~\ref{sec:package} provides an overview of the key features of the \texttt{R} package \texttt{shrinkDSM}. Section~\ref{sec:sim} presents the simulation study and evaluates the performance of the proposed method. Section~\ref{sec:appl} applies the method to a clinical dataset, illustrating its real-world utility. Finally, Section~\ref{sec:conc} concludes the paper.

\section{Model Specification and Estimation} \label{sec:model}

\subsection{The Normal Dynamic Survival Model} \label{sec:selection}

A defining characteristic of survival data is that it often experiences right censoring, which means that the failure time of an individual is not observed. Accordingly, there are two types of observations: those for which the failure has not yet occurred by the time of the last observation, and those where the last observation aligns with the failure time. An example for these two types of observations in an industrial context is one machine that was still observed to be running at the last inspection and another that failed at a previously observed point. To account for this, each individual $i = 1, \dots, N$ is assumed to have both a survival time $t_i$ and a censoring time $c_i$, which are independent random variables. The observed data of an individual then consists of the observed survival time $y_i = min(c_i, t_i)$, coupled with a failure indicator $d_i$, which is equal to $0$ if the observation is censored and $1$ otherwise, and a vector of $K$ covariates $\bm z_i = (z_{i1}, \dots, z_{iK})$. In our previous example the machine which had already failed would have $d_i = 1$ and $y_i = t_i$, while the one still running would have $d_j = 0$ and $y_j = c_j$. The survival times are measured relative to some origin point, such as the date of a medical procedure or the time point at which a machine is put into service. It is not required that the origin point be the same across observations. For example, if patients undergo surgery on different dates, the survival time of each patient would be measured from their date of surgery onwards.

The central quantity of interest in much of the survival literature is the \textit{hazard rate} $\lambda(t)$, i.e. the instantaneous rate of event occurrence
$$
\lambda(t)=\lim _{d t \rightarrow 0} \frac{P(t \leq t_i<t+d t \mid t_i \geq t)}{d t}.
$$
In the Cox proportional-hazards model, the hazard is modeled as
$$
\lambda\left(t | \bm{z}_{i}\right) = \exp \left(\beta_{0}(t)+\sum_{k=1}^{K} z_{i k} \beta_{k}\right),
$$
where only the baseline hazard $\beta_0(t)$ is allowed to vary over time, while the effect of the covariates on the hazard rate is assumed to be constant. \cite{gamerman1991dynamic} extended the Cox model by making both the baseline hazard $\beta_{0}$ as well as the effect of the covariates ($\beta_1(t), \dots, \beta_k(t)$) functions of time, i.e.
$$
\lambda\left(t | \bm{z}_{i}\right)=\exp \left(\beta_{0}(t)+\sum_{k=1}^{K} z_{i k} \beta_{k}(t)\right).
$$
The benefit of this extension is twofold: the first and obvious advantage is that the effect of the covariates can change throughout the observed survival time. A second, somewhat less obvious benefit, is that the model changes from a relative risk model to an absolute risk model, allowing for the estimation of a baseline hazard as well as the prediction of survival times given new covariate values.

As proposed by \cite{gamerman1991dynamic}, the dynamic survival model is a piecewise exponential model for lifetimes. The underlying assumption is that a given set $\mathcal{S}=\left\{s_{0}=0, s_{1}, \ldots, s_{J}\right\}, s_{0}<s_{1}<\cdots<s_{J}$ partitions the survival time axis into $J$ intervals $\left(s_{0}, s_{1}\right], \ldots,\left(s_{J-1}, s_{J}\right]$, relative to the individual origin points. Within an interval, the hazard is assumed to be constant. More formally, the baseline log-hazard and covariate effects $\beta_k(t)$, for $k = 0, \dots, K$, are defined by the piecewise constant functions
$\beta_k(t) = \beta_{kj}, \quad \text{for } t \in (s_{j-1}, s_j]$.
This, in turn, implies that the hazard also has a piecewise constant form:
\begin{equation} \label{eq:haz}
\lambda(t|\bm z_i; t \in (s_{j-1}, s_j]) = \lambda_{ij} = \exp\left(\beta_{0j} + \sum_{k=1}^{K} z_{i k} \beta_{kj}\right).
\end{equation}
In principle, as long as there is a value for each covariate, interval, and individual, the covariates can also vary over time, i.e.:
\begin{equation} \label{eq:haz_dyn}
\lambda(t|\bm z_{ij}; t \in (s_{j-1}, s_j]) = \lambda_{ij} = \exp\left(\beta_{0j} + \sum_{k=1}^{K} z_{ijk} \beta_{kj}\right).
\end{equation}
While the software introduced in section~\ref{sec:package} supports this, nothing fundamental about the model itself changes and therefore, to simplify notation, the interval index for the covariates will be omitted.

Since the partition of the time axis is fixed, its choice can significantly impact model fit, particularly if selected poorly. When a theoretical basis exists for a specific grid, it provides a natural choice. However, in many cases, no such basis is available. A simple solution is to use equally spaced intervals, but this can be problematic when event times are unevenly distributed, as the number of observations per interval may vary widely. To see why this can be an issue, consider a time interval that contains no observed events. In this case, the posterior is informed solely by the prior, as no data is available to provide direct information. A more sensible, data-driven approach is to place grid points whenever a fixed number of events have been observed (e.g., every two events). This ensures that all intervals contain the same amount of event-based information, improving stability and interpretability. This approach is implemented in the \texttt{R} package \texttt{shrinkDSM} via the \texttt{divisionpoints} function.

An important component of the model is the choice of stochastic process that governs the evolution of the $\beta_{kj}$'s from one time interval to the next. Following \cite{hemming2002parametric}, we consider the normal dynamic survival model, in which the covariate effects evolve according to Gaussian random walks, i.e.:
\begin{equation} \label{eq:rw}
\beta_{kj} = \beta_{k,j-1} + w_{kj}, \quad w_{kj} \sim \mathcal{N}\left(0, \theta_k\right),
\end{equation}
where the initial parameter value $\beta_{k0}$ is assumed to come from a normal distribution with mean $\beta_k$ and variance $\theta_k$, i.e. $\beta_{k0} \sim \mathcal{N}(\beta_k, \theta_k)$. The random walk specification assumes that the effect of covariates in neighboring intervals will not be vastly different from each other, thereby imposing a degree of smoothness. $\theta_k$, i.e. the variance of the innovations $w_{kj}$, plays a vital role in the degree of smoothness that is imposed through the random walk specification. Large variances allow for larger innovation values, thereby implying less similarity in the effect of the covariates on the hazard rate in neighboring intervals, while the inverse holds true for small variances.

\begin{figure}
  \centering
  \begin{knitrout}
  \definecolor{shadecolor}{rgb}{0.969, 0.969, 0.969}\color{fgcolor}
  \includegraphics[width=0.66\maxwidth]{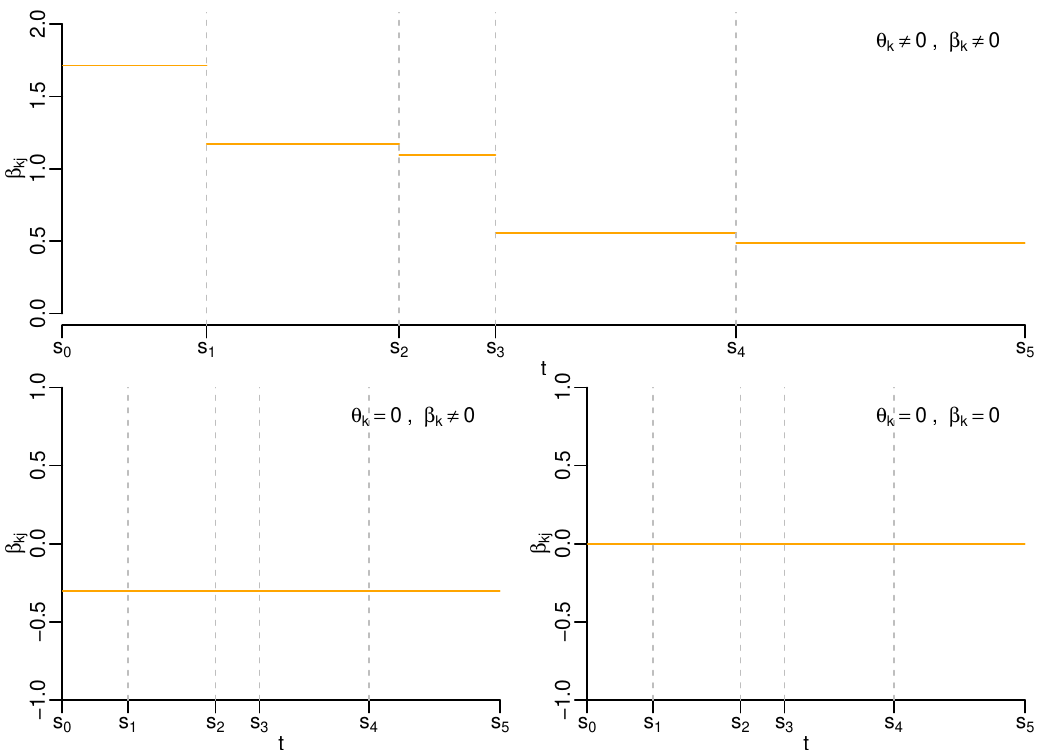}
  \end{knitrout}
  \caption{Exemplary evolution of the effect of three covariates on the hazard rate. The top figure is an example of a parameter $\beta_{kj}$ associated with a covariate that has a time-varying effect, while the bottom left and bottom right display $\beta_{kj}$'s where the covariate's effect is significant and non time-varying and insignificant and non time-varying, respectively.}
  \label{fig:betas}
\end{figure}

Taking this idea further, three interesting special cases arise out of the random walk specification. These can be differentiated based on whether the mean of the initial value $\beta_{j0}$ (i.e. $\beta_k$), and/or the variance of the innovations (i.e. $\theta_k$), are equal to zero. The baseline is the case where both $\beta_k \neq 0$ and $\theta_k \neq 0$. This implies that the covariate associated with this parameter already has an effect on the hazard at the origin and that this effect varies over time. The second case is when $\beta_k \neq 0$ but $\theta_k = 0$, resulting in a random walk where $\beta_{kj} = \beta_{k,j-1} = \dots = \beta_k$ with probability $1$. This means that the corresponding covariate has an effect on the hazard rate, however this effect remains constant at $\beta_k$ over time. The third and final interesting case is when both $\beta_k = 0$ and $\theta_k = 0$. In this case $\beta_{k0} = 0$ and all subsequent values $\beta_{kj} = \beta_{k0} = 0$ with probability $1$. Thus the covariate has no effect on the hazard and this does not change over time, effectively removing this covariate from the model. A visual example of each of these cases is shown in Figure~\ref{fig:betas}. Given a tool that can automatically differentiate between these cases leads to a setup that can choose which covariates to include and which to allow to vary over time. 

\subsection{Global-Local Shrinkage Priors for Variable and Variance Selection} \label{sec:priors}

Pulling $\beta_k$ and $\theta_k$ towards zero is a well studied problem in the realm of time-varying parameter models. A long string of literature uses various global-local shrinkage priors to effectively pull $\theta_k$ and $\beta_k$ towards zero where the data does not protest (see, e.g.,\cite{belmonte2014hierarchical, bitto2019achieving, cadonna2020triple}).  Despite their proven performance in differentiating between time-varying and constant parameters, this ability has not yet been introduced to the realm of dynamic survival models.

A general global-local shrinkage prior \citep{polson2011shrink} for a parameter $\beta_k$ takes the following form:
\begin{equation} \label{eq:gl_prior}
  \begin{gathered}
    \beta_k | \tau^\beta, \lambda^\beta_k  \sim \mathcal N \left(0, \lambda^\beta_k \tau^\beta \right), \quad \lambda^\beta_k \sim p_{\lambda^\beta}(\lambda^\beta_k) \\
    \tau^\beta \sim p_{\tau^\beta}(\tau^\beta),
  \end{gathered}
\end{equation}
where $\lambda^\beta_k$ represents the local parameter, while $\tau^\beta$ is the global shrinkage parameter. The intuition behind such priors is that the global parameter pulls all coefficients towards zero, while the local parameters allow individual parameters to escape this pull, where warranted by the data. Different choices for the distribution of $\lambda^\beta_k$ and $\tau^\beta$ determine the characteristics of the shrinkage prior, with common choices being the Bayesian Lasso \citep{park2008bayesian}, the normal-gamma prior \citep{brown2010inference} or the horseshoe prior \citep{carvalho2009handling}.

While this works well for pulling the mean of the initial value $\beta_k$ towards zero, the $\theta_k$'s are \textit{variance} parameters, and thus only live on $\mathbb R^+$. \cite{fruhwirth2010stochastic} adapted global-local shrinkage priors of the form \ref{eq:gl_prior} to variance parameters in the context of time-varying parameter models by introducing a non-centered parameterization, which led to a square transform version of global local shrinkage priors. If a random variable $X$ comes from the prior structure defined by \ref{eq:gl_prior} and we set $\theta_k = X^2$, then
\begin{equation} \label{eq:gl_prior_sq}
  \begin{gathered}
    \theta_k | \tau^\theta, \lambda^\theta_k  \sim \mathcal G \left(\frac 1 2, \frac 1 {2\lambda^\theta_k \tau^\theta} \right), \quad \lambda^\theta_k \sim p_\lambda^\theta(\lambda^\theta_k) \\
    \tau^\theta \sim p_{\tau^\theta}(\tau^\theta).
  \end{gathered}
\end{equation}
This defines a class of global-local shrinkage priors specifically for variance parameters, depending on the choice of distribution for $\lambda^\theta_k$ and $\tau^\theta$. 

Our choices for $\lambda_k^\beta$, $\lambda_k^\theta$, $\tau^\beta$ and $\tau^\theta$, follow \cite{cadonna2020triple} and employ the triple gamma shrinkage prior, which can be represented as
\begin{equation} \label{eq:tg_prior}
  \begin{gathered}
    \theta_k | \tau^\theta, \lambda_k^\theta  \sim \mathcal G \left(\frac 1 2, \frac 1 {2\lambda_k^\theta \tau^\theta} \right), \quad \lambda_k^\theta \sim F\left(2a^\theta, 2c^\theta\right) \\
    \tau^\theta \sim F\left(2c^\theta, 2a^\theta\right),
  \end{gathered}
\end{equation}
where $F(d_1, d_2)$ is the F-distribution with $d_1$ and $d_2$ degrees of freedom. Similarly, we place the variable selection twin of the triple gamma prior, the normal-gamma-gamma prior \citep{griffin2017hierarchical}, on the mean of the initial value $\beta_k$:
\begin{equation} \label{eq:ngg_prior}
  \begin{gathered}
    \beta_k | \tau^\beta, \lambda_k^\beta  \sim \mathcal N \left(0, \lambda_k^\beta \tau^\beta \right), \quad \lambda_k^\beta \sim F\left(2a^\beta, 2c^\beta\right) \\
    \tau^\beta \sim F\left(2c^\beta, 2a^\beta\right).
  \end{gathered}
\end{equation}
The prior can be seen as a strong choice for two main reasons. First, it allows for fairly granular control over the shrinkage characteristics of the prior, as the parameters $a^\theta$ ($a^\beta$) and $c^\theta$ ($c^\beta$) explicitly control behavior of the prior around zero and in the tails, respectively. More concretely, smaller values of $a^\theta$ ($a^\beta$) lead to a prior that places more mass around 0, leading to heavier shrinkage of smaller parameter values. In particular, values of $a^\theta$ ($a^\beta$) smaller than $0.5$ lead to prior that is unbounded at $0$, implying strong shrinkage around the origin. Conversely, smaller values of $c^\theta$ ($c^\beta$) lead to a prior that places more mass in the tails, thereby incurring less of a shrinkage penalty for large parameter values. These two hyperparameters in tandem can therefore ensure that noise is properly squelched, while signal can filter through. Second, it can be considered a fairly generic prior, as it subsumes many popular prior choices, such as the horseshoe \citep{carvalho2009handling}, the Bayesian Lasso \citep{park2008bayesian}, the normal-gamma \citep{brown2010inference} and a simple Gaussian prior. Interested readers are referred to \cite{cadonna2020triple} for a more in-depth discussion, in particular due to the focus on Gaussian state space models. 

Placing priors on the global parameters $a^{\theta}$, $c^{\theta}$, $a^{\beta}$ and $c^{\beta}$ allows them to be learned from the data and gives the model the ability to adapt to the level of sparsity in any given dataset. Here, we follow \cite{cadonna2020triple} and set scaled beta distributions as priors:
$$
\begin{aligned}
2 a^{\theta} \sim \mathcal{B}\left(\alpha_{a^\theta}, \beta_{a^\theta}\right), & \quad 2 c^{\theta} \sim \mathcal{B}\left(\alpha_{c^\theta}, \beta_{c^\theta}\right), \\
2 a^{\beta} \sim \mathcal{B}\left(\alpha_{a^{\beta}}, \beta_{a^{\beta}}\right), & \quad 2 c^{\beta} \sim \mathcal{B}\left(\alpha_{c^{\beta}}, \beta_{c^{\beta}}\right).
\end{aligned}
$$
These distributions force the parameters to stay in the range $(0, 0.5)$. As discussed above, this leads to a prior that induces heavy shrinkage near $0$, while still allowing for heavy enough tails to let signals filter through.

\subsection{Estimation Procedure} \label{sec:estim}

Estimation of the unknown parameters is done through MCMC. We extend the approach proposed by \cite{wagner2011bayesian}, which uses data augmentation to enable Gibbs sampling and avoid Metropolis-Hastings (MH) steps where possible. While this is feasible for most parameters, the conditional posteriors of $a^{\theta}$, $c^{\theta}$, $a^{\beta}$ and $c^{\beta}$ are not of well-known form and therefore require MH-within-Gibbs steps. The sampler also leans on ideas from \cite{bitto2019achieving} and \cite{cadonna2020triple} for sampling steps of parameters associated with the shrinkage priors. The core idea behind the sampler is to use data augmentation to cast the model into a conditionally Gaussian state space form, a class of models for which a wealth of efficient sampling algorithms exist. As such, we only elaborate on the data augmentation scheme in the following section and refer the interested reader to the supplementary material for the full MCMC algorithm.

\subsubsection{Data Augmentation} \label{sec:dataug}

A key ingredient of the MCMC sampler is the data augmentation scheme introduced by \cite{wagner2011bayesian}. She shows that the model can be cast into a conditionally Gaussian state space form through the use of a mixture approximation of the standard Gumbel distribution (also known as the generalized extreme value distribution type-I). Beyond simply avoiding MH steps, this allows the use of one-block samplers for the $\beta_{kj}$'s with tools such as forward-filtering, backward-sampling \citep{fruewirth1994data, carter1994on} and related algorithms (e.g., \cite{mccausland2011simulation}).
We define $u_{ij}$ as the survival time of individual $i$ within the time interval $j = (s_{j-1}, s_j]$. Each individual for which failure is not observed by the end of the interval is considered right censored, with the censoring point at $s_j$. To capture the censoring, a residual survival time $\xi_{ij}$ is added to each censored observation $u_{ij}$. This gives rise to the total survival time $\tau_{ij} = u_{ij} + \xi_{ij}$ of individual $i$ in period $j$. Due to the memorylessness of the exponential distribution, the residual survival time also follows an exponential $\text{Ex}(\lambda_{ij})$ distribution, conditional on $\tau_{ij} > u_{ij}$. Thus one can write these auxiliary survival times as
$$
\tau_{ij} = u_{ij} + \xi_{ij}, \quad \xi_{ij} \sim \text{Ex}(\lambda_{ij}) \text{ for } j = 1, \dots, l_i-1,
$$
with the form of the final auxiliary survival time depending again on the indicator $d_i$:
$$
\tau_{il_i} =
\begin{cases}
u_{il_i} & \text{if } d_i = 1, \\
u_{il_i} + \xi_{il_i}, \quad \xi_{il_i} \sim \text{Ex}(\lambda_{il_i}) & \text{if } d_i = 0.
\end{cases}
$$

Represented this way, the normal dynamic survival model can be viewed as a generalized dynamic linear model, where the observations are conditionally exponentially distributed with the hazard rate $\lambda_{ij}$ acting as the rate parameter:
$$
\begin{array}{l} \label{eq:gdlm}
\tau_{i j} | \bm{\beta}_{j} \sim \operatorname{Ex}\left(\exp \left(\bm{z}_{i} \bm{\beta}_{j}\right)\right), \\
\bm{\beta}_{j}=\bm{\beta}_{j-1}+\bm{\omega}_{j}, \quad \bm{\omega}_{j} \sim N(\bm{0}, \bm{Q}(\bm{\theta})).
\end{array}
$$
In this notation, the baseline hazard is absorbed into $\bm \beta_j = (\beta_{0j}, \beta_{1j}, \dots, \beta_{Kj})'$, implying that an intercept term is added to $\bm z_i = (1, z_{i1}, \dots, z_{iK})$. The variances of the innovations are collected in the diagonal matrix $\bm Q(\theta) = \text{diag }(\theta_0, \theta_1, \dots, \theta_K)$, which also makes the assumption that the random walks evolve independently of one another explicit.
Taking the log of the observation equation transforms it to be linear:
$-\ln \tau_{i j}=\bm{z}_{i} \bm{\beta}_{j}+\epsilon_{i j}$,
where $\epsilon_{i j}$ now follows a standard Gumbel distribution. \cite{fruhwirth2007auxiliary} show that this distribution can be approximated very closely through a mixture of 10 normal distributions
$$
p_{\epsilon}(\epsilon)=\exp \left(-\epsilon- \exp(-\epsilon)\right) \approx \sum_{r=1}^{10} w_{r} f_{\mathrm{N}}\left(\epsilon ; m_{r}, v_{r}\right).
$$
The parameters of the mixture distribution, i.e. the weights $w_r$, the means $m_r$ and the variances $v_r$, are determined by minimizing the Kullback-Leibler divergence from the approximating mixture to the Gumbel distribution. The exact values can be found in Table 1 in \cite{fruhwirth2007auxiliary}.

After introducing the component indicators $r_{i j} \in\{1, \ldots, 10\}$, the model can be written in conditionally Gaussian state space form:
$$
-\ln \tau_{i j}=\bm{z}_{i} \bm{\beta}_{j}+m_{r_{i j}}+\varepsilon_{r_{i j}}, \quad \varepsilon_{r_{i j}} \sim N\left(0, v_{r_{i j}}\right).
$$
This concept can be illustrated more clearly by consolidating all observations still at risk, i.e., those that have not yet exited due to failure or censoring, at the beginning of a given interval into a single equation. To this end, let $n_j$ be all observations still at risk in a given interval $j$. Furthermore, assume that the vector of observed survival times $\bm y$ is sorted from longest to shortest, with $y_1$ being the longest and $y_N$ being the shortest, and that the rows of the design matrix $\bm Z$ were ordered with the same key. Then, through defining a latent vector $\bm x_j$ as
$$
\bm{x}_{j} =
\begin{pmatrix}
-\ln \tau_{1 j}-m_{r_{1}, j} \\
\vdots \\
-\ln \tau_{n_{j}, j}-m_{r_{n_{j}}, j}
\end{pmatrix}
$$
and $\bm\varepsilon_{j}=\left(\varepsilon_{r_{1}, j}, \ldots, \varepsilon_{r_{n_{j}}, j}\right)^\prime$, the model can be written in the following conditionally Gaussian state space form
\begin{equation} \label{eq:gssm}
\begin{aligned}
\bm{x}_{j}&=\bm{Z}_{j} \bm{\beta}_{j}+\bm{\varepsilon}_{j}, \quad \bm{\varepsilon}_{j} \sim N\left(\bm{0}, \bm{V}_{j}\right), \\
\bm{\beta}_{j}&=\bm{\beta}_{j-1}+\bm{\omega}_{j}, \quad \bm{\omega}_{j} \sim N(\bm{0}, \bm{Q}(\bm{\theta})),
\end{aligned}
\end{equation}
where $\bm V_j = \text{diag}\left(v_{r_{1 j}}, \ldots, v_{r_{n_{j}, j}}\right)$ and $\bm Z_j$ consists of the first $n_j$ rows of the reordered design matrix, including the intercept column:
$$
\bm Z_j =
\begin{pmatrix}
\bm z_1 \\
\vdots \\
\bm{z}_{n_j}
\end{pmatrix}.
$$

Thus, using these data augmentation steps, it is possible to represent the normal dynamic survival model defined in \eqref{eq:haz} and \eqref{eq:rw} as a conditionally Gaussian state space model \eqref{eq:gssm}, opening up the possibility to use sampling procedures from this well understood class of models. The interested reader is referred to the supplementary material for the full MCMC algorithm.

\subsection{Extension to a Time-Varying Grouped Factor Model}\label{sec:factor}
To account for unobserved heterogeneity that may be present in the augmented survival times in the form of covariance, we additionally add a grouped factor component to the hazard rates. Let observation $i$ belong to group $g$, with $g \in \{1, \dots, G\}$. Note that this grouping is assumed to be observed and is not estimated from the data. Then, when the factor component is added, the hazard rates look as follows:

$$
\lambda_{ij} = \exp\left(\phi_g f_j + \beta_{0j} + \sum_{k=1}^{K} z_{i k} \beta_{kj}\right),
$$

where $f_j$ is a single latent factor which is allowed to vary over time according to a zero-mean stochastic volatility law of motion:

$$
\begin{aligned}
f_{j} \mid h_{j} & \sim \mathcal{N}\left(0, \exp(h_{j})\right), \\
h_{j} \mid h_{j-1}, \phi_{f}, \sigma_{f}^{2} & \sim \mathcal{N}\left(\phi_{f}h_{j-1}, \sigma_{f}^{2}\right), \\
h_{0} \mid \phi_{f}, \sigma_{f}^{2} & \sim \mathcal{N}\left(0, \sigma_{f}^{2} /\left(1-\phi_{f}^{2}\right)\right).
\end{aligned}
$$

This specification has similarities with shared frailty models \cite{hougaard2000shared}, with the special case $G = N$ closely related to regular (i.e. individual level) frailty models \cite{hougaard1995frailty}. A key difference lies in the stochastic volatility specification, as it allows for time variation in the frailty term, whereas standard frailty models usually assume that the term remains constant over time. Frailty models are random effect models applied in the context of time-to-event analysis, where the aim is to capture unobserved heterogeneity. Such unobserved heterogeneity, where it impacts the outcome variable, can induce spurious time-dependent effects of covariates through a selection effect \cite{balan2020tutorial}. This effect also has ramifications for the identifiability of frailty terms in models with time-varying parameters, as, without strict assumptions, it is not possible to determine whether time-variation comes from individual frailty or through a truly time-varying effect. In the context of a time-varying parameter model, frailty is therefore particularly useful in the shared frailty specification, where multiple observations share a single frailty term, as this alleviates the identifiability issues. A common example of an application for such a shared frailty model is a scenario where repeated measurements are taken, as the shared frailty term can account for unobserved individual heterogeneity.

In this specification, the signs of the factor loadings $\phi_1, \dots, \phi_G$ and the factor $f_j$ are not identified. To demonstrate this, let $\bm \phi$ be the vector of factor loadings for all individuals. Then
$$
\bm \phi f_j = (-\bm \phi)(-f_j).
$$
To remedy this, we first run the MCMC sampler in the unrestricted space and then enforce identification post-hoc through the methods described in \cite{fruhwirth2024sparse} \cite{lopes2004bayesian}. In the one-dimensional case (as it appears above), this amounts to ensuring that $\phi_1$ is always positive.

In specifying the priors for the parameters governing the stochastic volatility process, we follow \cite{kastner2016dealing}:

$$
\begin{aligned}
\left(\phi_{f}+1\right) / 2 \mid a_{0}, b_{0} &\sim \mathcal{B}\left(a_{0}, b_{0}\right), \\
\sigma_{f}^{2} \mid B_{\sigma_{f}} &\sim \mathcal{G}\left(1 / 2,1 / 2 B_{\sigma_{f}}\right).
\end{aligned}
$$
To ensure sparsity in the factor loadings, we place triple gamma shrinkage priors on the factor loadings, analogous to those used in Section~\ref{sec:priors}:
$$
\phi_{g} \mid a_\phi, c_\phi, \lambda^2_{B\phi} \sim TG(a_\phi, c_\phi, \lambda^2_{B\phi}).
$$
Naturally, as in Section~\ref{sec:priors}, this includes all special and limiting cases of the triple gamma prior, allowing for a large degree in flexibility in terms of the amount of shrinkage applied. Similarly, the hyperpriors are also specified in an analogous fashion to those in Section~\ref{sec:priors}. This approach has similarities to that proposed by \cite{kastner2019sparse}, albeit with a more general choice of shrinkage prior.
This estimation can be extended straightforwardly to account for the additional factor component, thanks to the flexible nature of MCMC. The additional steps required are described in detail in Algorithm 2 in the Appendix.

\section{\texttt{R} Package shrinkDSM}
\label{sec:package}
The methods described are implemented in the R  package \texttt{shrinkDSM}\cite{winkler2022shrinkdsm}, offering an efficient yet easy-to-use tool for estimating piecewise exponential dynamic survival models. It aims to have a low barrier to entry, while still being flexible enough for more advanced users. This is achieved by choosing sensible default values for the hyperparameters, while allowing the user to modify the prior setup in fairly deep ways, should they so desire. The resulting software works well for a wide variety of datasets, requires very little tuning and leads to a default model that can be estimated in only a few lines:
\begin{knitrout}
  \definecolor{shadecolor}{rgb}{0.969, 0.969, 0.969}\color{fgcolor}
  \begin{kframe}
  \setstretch{1.0}
  \begin{alltt}
  \hlkwd{set.seed}\hlstd{(}\hlnum{123}\hlstd{)}
  \hlkwd{data}\hlstd{(}\hlstr{"gastric"}\hlstd{)}
  
  \hlcom{# Create intervals for piecewise}
  \hlcom{# exponential model}
  \hlstd{intvls} \hlkwb{<-} \hlkwd{divisionpoints}\hlstd{(gastric}\hlopt{$}\hlstd{time,} 
                           \hlstd{gastric}\hlopt{$}\hlstd{status,} 
                           \hlnum{2}\hlstd{)}
  
  \hlcom{# Estimate default model}
  \hlstd{mod} \hlkwb{<-} \hlkwd{shrinkDSM}\hlstd{(time} \hlopt{~} \hlstd{radiation,}
                   \hlkwc{gastric,}
                   \hlkwc{delta} \hlstd{= gastric}\hlopt{$}\hlstd{status,} 
                   \hlkwc{S} \hlstd{= intvls)}
  \end{alltt}
  \end{kframe}
  \end{knitrout}
  
More advanced users can modify the prior setup (e.g. change the class of prior used, modify hyperparameters or even deactivate the learning of certain hyperparameters) in a fashion analogous to the methods presented by \cite{knaus2021shrinkage}, as the underlying Gaussian state space model is very similar. While this similarity leads to many of the methods carrying over, a distinction lies in the factor model, which is only present in the model described in this paper. To add a factor to the model, a grouping variable is supplied via the argument \texttt{groups}. Then, the prior setup for $\phi_1, \dots, \phi_G$ can be modified via the argument \texttt{phi\_param}:
\begin{knitrout}
\definecolor{shadecolor}{rgb}{0.969, 0.969, 0.969}\color{fgcolor}
\begin{kframe}
\setstretch{1.0}
\begin{alltt}
\hlcom{# Modify hyperparameters for phi}
\hlcom{# Note that using group = 1:nrow(gastric)}
\hlcom{# is the special case G = N}
\hlstd{mod2} \hlkwb{<-} \hlkwd{shrinkDSM}\hlstd{(time} \hlopt{~} \hlstd{radiation,}
                  \hlstd{gastric,}
                  \hlkwc{delta} \hlstd{= gastric}\hlopt{$}\hlstd{status,} 
                  \hlkwc{S} \hlstd{= intvls,}
                  \hlkwc{group} \hlstd{=} \hlnum{1}\hlopt{:}\hlkwd{nrow}\hlstd{(gastric),}
                  \hlkwc{phi_param} \hlstd{=} \hlkwd{list}\hlstd{(}
                   \hlkwc{learn_a_phi} \hlstd{=} \hlstr{FALSE}\hlstd{,}
                   \hlkwc{a_phi} \hlstd{=} \hlnum{0.1}
                  \hlstd{))}
\end{alltt}
\end{kframe}
\end{knitrout}
The arguments provided in the named list for \texttt{phi\_param} behave in the same fashion as those for the variances of the innovations and the means of the initial states, albeit with the suffix \texttt{\_phi}.

A further goal is to combine the benefits of the low-level (and therefore fast) programming language \texttt{C++} with those of the higher level language \texttt{R}, with its interpreted code and rich suite of data wrangling software. All computationally intensive code is implemented in \texttt{C++} and then interfaced with \texttt{R} via the \texttt{Rcpp}\cite{eddelbuettel2011rcpp} package. Code dealing with plotting and interpreting results, as well as checking MCMC convergence, is then written in \texttt{R}, allowing for easy interfacing with other \texttt{R} packages (e.g. \texttt{coda}\cite{plummer2006coda}). Furthermore, forecasting of survival times based on new covariates and the use of time-varying covariates is supported.

\section{Simulation Study} \label{sec:sim}

\begin{figure*}
  \includegraphics[width=\textwidth]{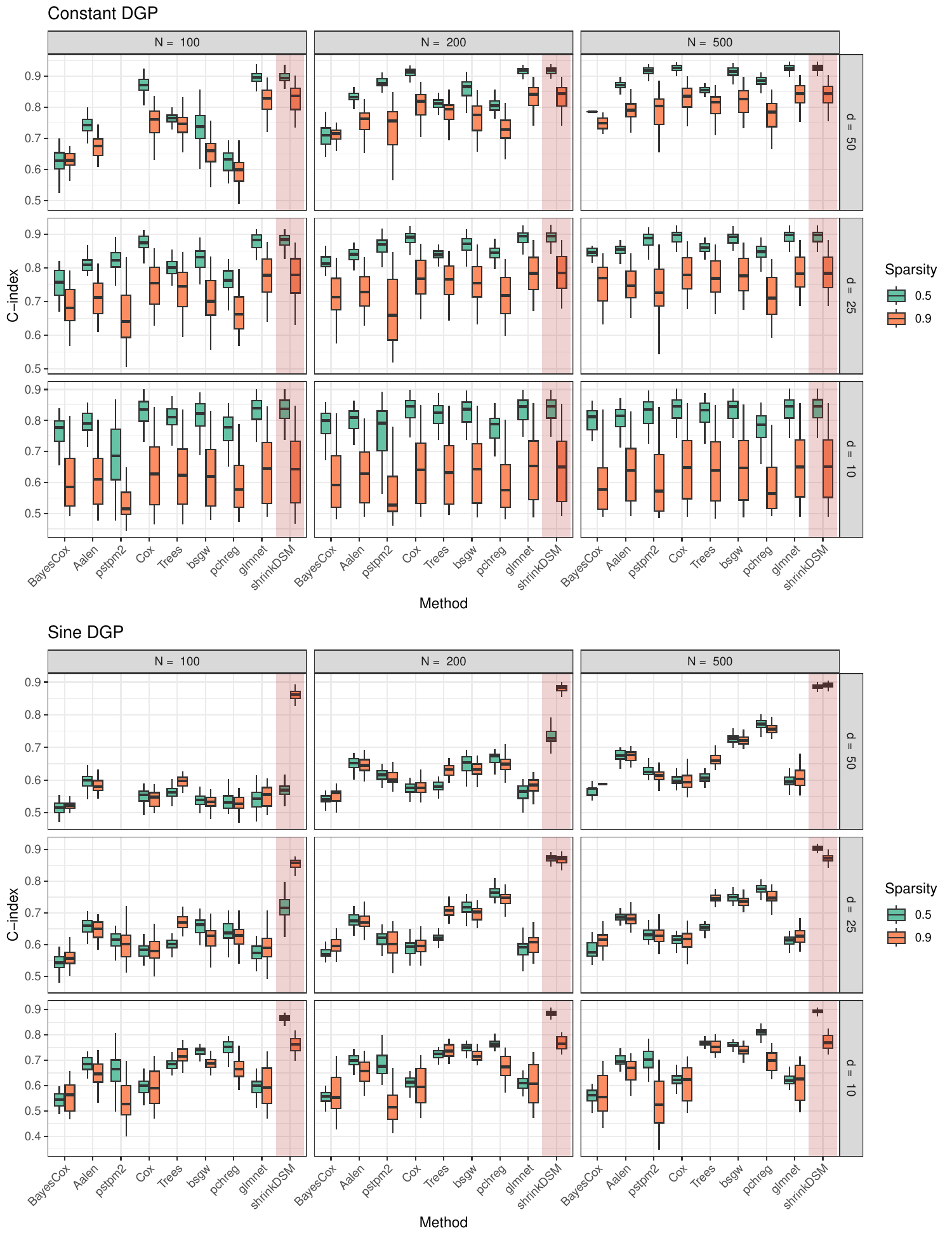}
  \caption{Distribution of concordance index over 50 repetitions under the constant and sinusoidal DGP. Models are ordered by the median concordance index across all simulations. The red overlay highlights the proposed method.}
  \label{fig:sim_results_const_sin}
\end{figure*}

\begin{figure*}
  \includegraphics[width=\textwidth]{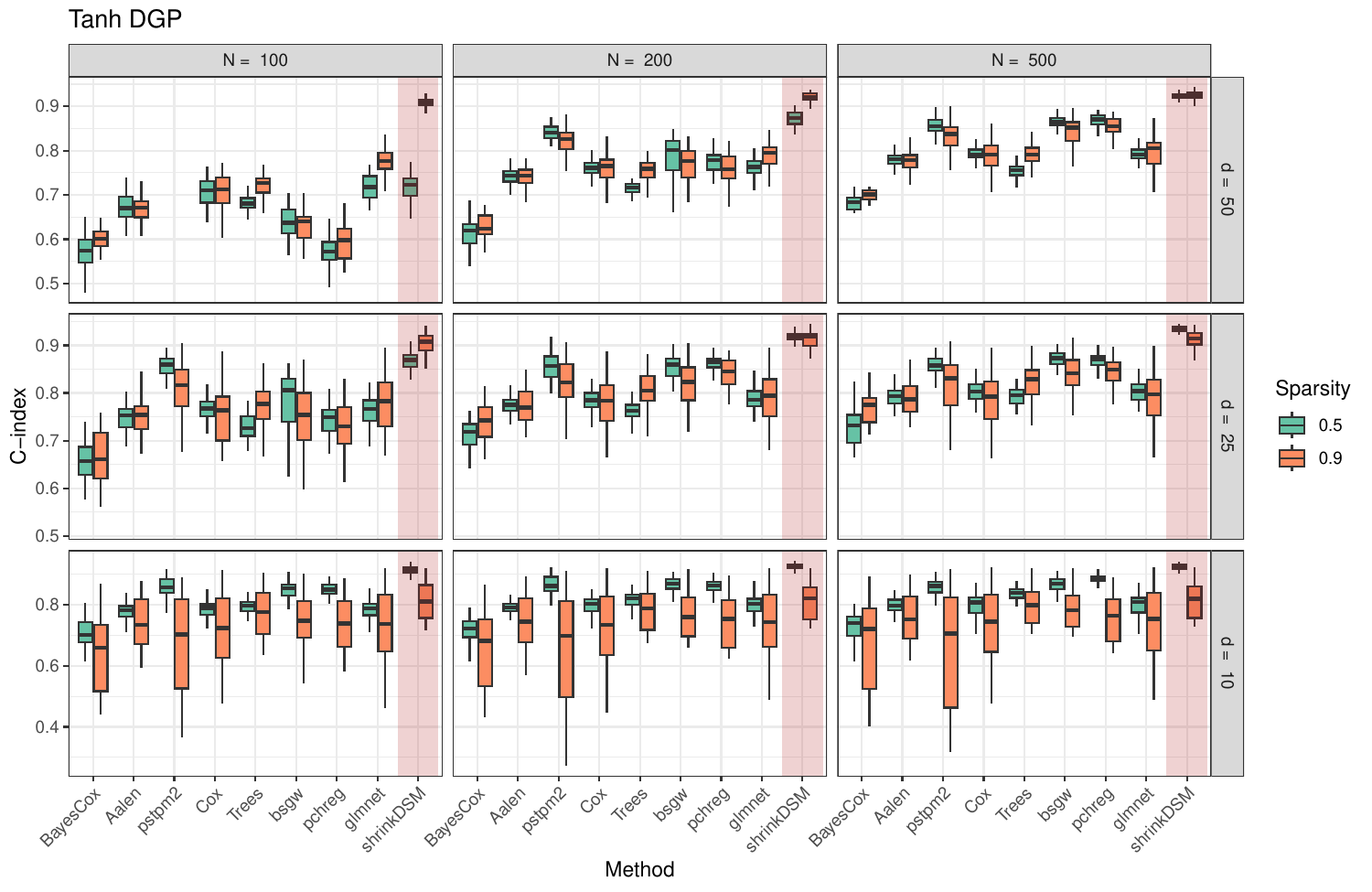}
  \caption{Distribution of concordance index over 50 repetitions under the tanh DGP. Models are ordered by the median concordance index across all simulations. The red overlay highlights the proposed method.}
 \label{fig:sim_results_tanh}
\end{figure*}

\begin{table*}[]
  \renewcommand{\arraystretch}{1.2} 
  \tiny
  \begin{tabular}{@{}lllll@{}}
  \toprule
  R package & R function & Function settings (if different from default) & Penalized & Time-varying\\ \midrule
  \texttt{dynsurv} \citep{dynsurv-package} & \texttt{bayesCox} & \makecell[l]{Grid set at every third observed event, \\ default AR(1) prior for coefficients.} & X & X\\
  \texttt{timereg} \citep{scheike2011analyzing} & \texttt{aalen} & 1000 simulations used for resampling. &  & X \\
  \texttt{rstpm2} \citep{xing2018parametric} & \texttt{pstpm2} & All coefficients estimated as penalized splines. & X & X \\ 
  \texttt{BSGW} \citep{bsgw} & \texttt{bsgw} & Function defaults. & X & \\
  \texttt{survival} \citep{survival-package} & \texttt{coxph} & Maximum 1000 iterations for convergence. & &  \\
  \texttt{pch} \citep{pch} & \texttt{pchreg} & Function defaults. & & X\\
  \texttt{ranger} \citep{wright2017ranger} & \texttt{ranger} & \makecell[l]{\texttt{splitrule} set to \texttt{extratrees}, \texttt{mtry} set to 4, \\ \texttt{importance} set to \texttt{permutation}.} & (X) & \\
  \texttt{glmnet} \citep{friedman2010regularization, simon2011regularization} & \texttt{cv.glmnet}  & \makecell[l]{Cross-validation measure set to \\ concordance index. Lasso penalty (\(\alpha = 1\)) \\ used to enforce sparsity.} & X & \\ \bottomrule                  
  \end{tabular}
  \
  \caption{Overview of R packages and functions used in the simulation study.}
  \label{tab:sim_packages}
\end{table*}

To evaluate the performance of the proposed model, we benchmark it against a set of competing methods in a simulation study.

\subsection{Simulation Setup and Competing Methods}
\begin{figure}
  \centering
  \includegraphics[width=0.66\maxwidth]{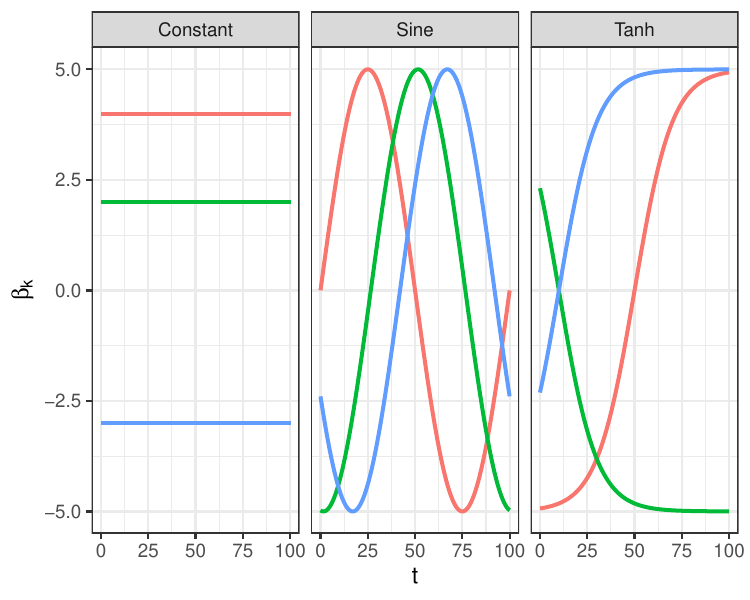}
  \caption{Examples of coefficient trajectories over time for different DGPs in the simulation study. Each panel represents a different DGP: constant, sinusoidal, and hyperbolic tangent.}
  \label{fig:sim_dgps}
\end{figure}

The synthetic data is generated under three different data-generating processes (DGPs), each differing in the functional form of the regression coefficients:

\begin{itemize}
  \item \textit{Constant DGP}: The coefficients remain constant over time, corresponding to the standard Cox proportional hazards model. 
  \item \textit{Sinusoidal DGP}: The coefficients follow a sinusoidal function of continuous time:
  \[
  \beta_{k}(t) = 5\sin\left(\frac{2\pi}{100}t - s_k\right),
  \]
  where \( s_k \) is a phase shift drawn from a uniform distribution over \( [0, 1.6\pi] \). Under this specification, each coefficient completes a full oscillation every 100 time units.
  \item \textit{Tanh DGP}: The coefficients follow a hyperbolic tangent function:
  \[
  \beta_{k}(t) = d_k \cdot 5 \tanh\left(\frac{5}{100}(t - i_k)\right),
  \]
  where \( d_k \in \{-1, 1\} \) determines whether the transition is increasing (\( d_k = 1 \)) or decreasing (\( d_k = -1 \)), and \( i_k \) is the inflection point, randomly drawn from \( [0, 100] \).
\end{itemize}
Examples of coefficient trajectories under these DGPs are shown in Figure~\ref{fig:sim_dgps}. 

Beyond the different DGPs, the simulations vary along the following dimensions:
\begin{itemize}
    \item \textbf{Sparsity levels}: \( s \in \{0.5, 0.9\} \).
    \item \textbf{Dimensionality}: The number of covariates, \( K \in \{10, 25, 50\} \).
    \item \textbf{Sample sizes}: The number of observations, \( N \in \{100, 200, 500\} \).
\end{itemize}
The covariates are drawn from a standard normal distribution, i.e., \( z_{ik} \sim \mathcal{N}(0,1) \). To introduce sparsity, a fraction \( s \) of coefficients is set to zero, meaning \(\lfloor s \cdot K \rfloor\) coefficients are randomly zeroed out in each simulation repetition.

Survival times are generated in three steps:

\begin{enumerate}
    \item \textbf{Time Discretization}: The time axis from \( t = 0 \) to \( t = 100 \) is divided into 5000 equally spaced intervals of size \( 0.02 \). This ensures a fine-grained approximation of continuous-time survival models while remaining computationally efficient.

    \item \textbf{Hazard Calculation}: At each time step, the covariate-driven hazard multiplier is computed as
    \[
    \lambda_{\text{cov}} = \exp\left(\sum_{k=1}^K z_{ik} \beta_{kj}\right).
    \]
    The baseline hazard is dynamically scaled at each step to normalize for covariate effects, ensuring that the overall hazard remains comparable across DGPs and results in a censoring proportion of approximately 30\%.

    \item \textbf{Survival Time Simulation}: Given the hazard function, survival times are drawn from a piecewise exponential distribution, assuming the hazard remains constant within each interval. Due to the high resolution of the time grid, this closely approximates a continuous-time survival model.
\end{enumerate}

Each combination of DGP, sparsity, dimensionality and sample size is repeated 50 times. Model performance is evaluated using the concordance index (C-index), calculated on 1000 out-of-sample observations. For models with time-varying coefficients, risk scores are computed as the negative mean survival time, averaged over 10,000 posterior predictions. This ensures that risk scores reflect the expected survival trajectory while remaining comparable across models. For models with constant coefficients, the C-index is computed based on their respective estimated risk scores. The C-index measures the proportion of concordant observation pairs, i.e., pairs where the individual with the shorter survival time also has a higher risk score. An overview of the competing methods can be found in Table~\ref{tab:sim_packages}.

The proposed model is estimated using a Markov chain Monte Carlo (MCMC) algorithm with 50,000 iterations, a burn-in of 10,000, and thinning of 10. The piecewise exponential model grid is set at every second observed event. The prior setup is as described in Section~\ref{sec:priors}, with the hyperparameters for the shrinkage priors set to  
\[
\alpha_{a^\beta} = \alpha_{c^\beta} = 5, \quad \beta_{a^\theta} = \beta_{c^\theta} = 10.
\]
This choice of hyperparameters favors smaller values of \(a^\beta, a^\theta, c^\beta\), and \(c^\theta\), leading to a shrinkage prior with a highly pronounced peak at zero and heavy tails, which is particularly useful in high-dimensional settings. 

\subsection{Results}

The simulation results are presented in Figure~\ref{fig:sim_results_const_sin} and Figure~\ref{fig:sim_results_tanh}.

The proposed method performs consistently well across all scenarios, either outperforming or remaining competitive with alternative approaches. Its advantage is particularly pronounced in settings where the true data-generating process (DGP) involves time-varying coefficients, as seen in the sinusoidal and tanh DGPs. Interestingly, competing methods that explicitly model time variation (\texttt{bayesCox}, \texttt{aalen}, \texttt{pstpm2}, \texttt{pchreg}) do not consistently outperform methods that assume constant coefficients, even when penalization is applied (\texttt{pstpm2}). This suggests that standard approaches to incorporating time variation may not be sufficient in high-dimensional or sparse settings, where adaptive shrinkage is crucial to prevent overfitting. This strong performance of the proposed method in time-varying settings is expected. More notably, however, it remains competitive in the constant DGP, demonstrating that the hierarchical shrinkage prior effectively suppresses unnecessary flexibility when time-varying effects are absent. This suggests that the model successfully balances adaptivity and regularization, avoiding overfitting in cases where a simpler model would suffice. Moreover, even when the underlying coefficients are truly constant, the triple gamma prior effectively performs variable selection, as demonstrated by the strong performance in high-dimensional and highly sparse settings. This indicates that the proposed approach is not just a flexible alternative to existing models but a robust method that adapts to the structure of the data without excessive tuning. Overall, the proposed method hedges against model misspecification, performing well regardless of whether the true underlying DGP exhibits time variation. This adaptability makes the method particularly valuable in real-world applications, where the presence of time variation is unknown and misspecification risks are high.

\section{Application to a Data Set of Patients with Adenocarcinoma of the Gastroesophageal Junction} \label{sec:appl}

\begin{table}[hb]
  \centering
  \caption{Description of explanatory variables used in application}
  \label{tab:variables}
\begin{tabular}{@{}ll@{}}
\toprule
Variable & Description                                                          \\ \midrule
SII      & Systemic Immune Inflammation Index                                    \\
age      & Age at the beginning of treatment                                    \\
sex      & Female = 0, male = 1                                                 \\
cT       & Clinical tumor stage                                                 \\
cN       & Clinical lymph node stage                                            \\
ASA      & \makecell[l]{American Society of Anesthesiologists \\ physical status classification}  \\
G        & Tumor differentiation                                                \\ \bottomrule
\end{tabular}
\end{table}

\begin{figure*}[htbp]
  \centering
  \includegraphics[width=\textwidth]{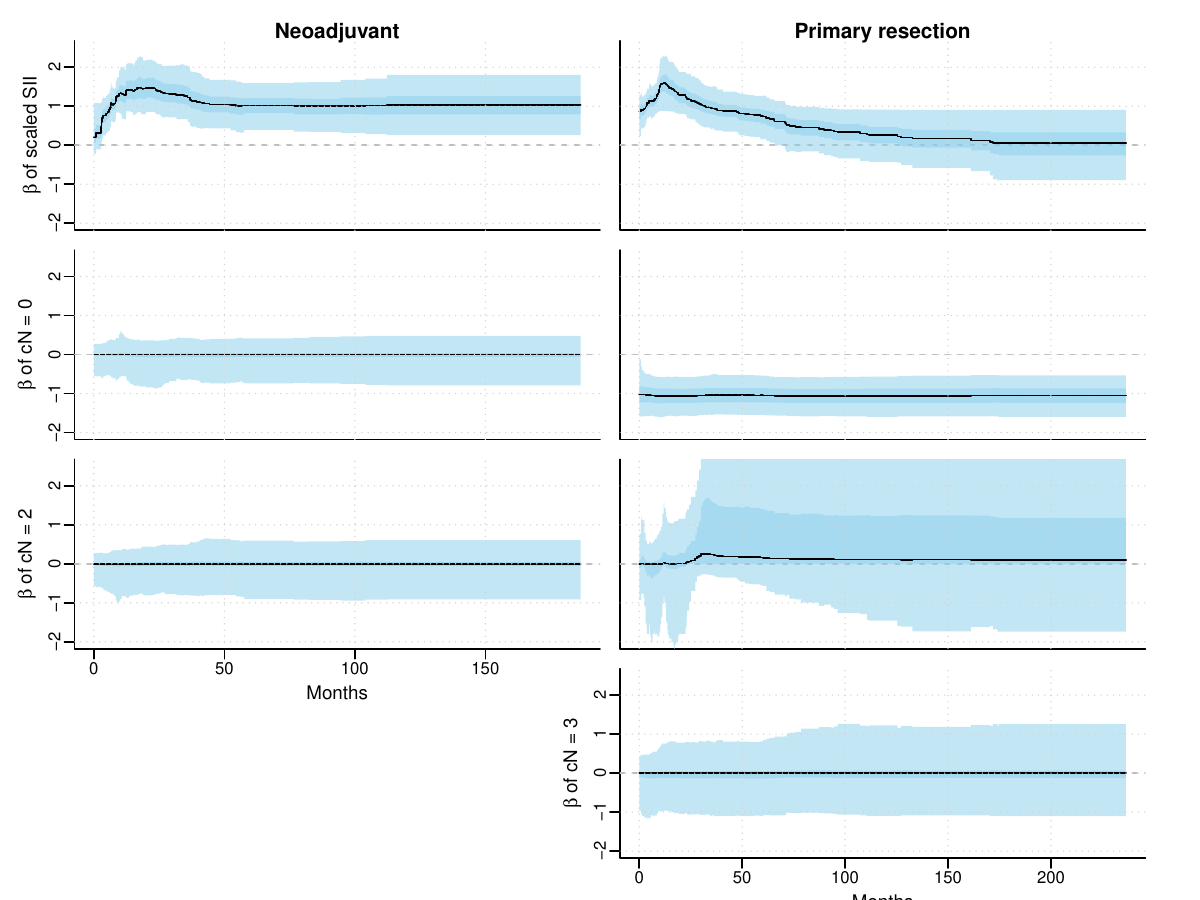}
  \caption{Posterior parameters associated with systemic immune-inﬂammation index (SII) and clinical lymphnode stage (cN) for primary resected and neoadjuvant treated patients over time. Black lines represent posterior medians, while shaded areas represent 95\% and 50\% pointwise credible intervals.}
  \label{fig:posteriors}
\end{figure*}

\begin{table*}[htbp]
  \centering
  \setstretch{1.0}
  \caption{Parameters estimated using the Cox proportional-hazards model}
  \label{tab:coxph}
\begin{tabular}{@{\extracolsep{5pt}}lcc}
\\[-1.8ex]\hline
\hline \\[-1.8ex]
 & \multicolumn{2}{c}{\textit{Dependent variable:}} \\
\cmidrule{2-3}
\\[-1.8ex] & \multicolumn{2}{c}{Survival time} \\
\\[-1.8ex] & (Neoadjuvant) & (Primary resection)\\
\hline \\[-1.8ex]
 scaled SII & 1.250$^{***}$ & 1.024$^{***}$ \\
  & (0.137) & (0.104) \\
  cT2 & 0.203 &  \\
  & (0.248) &  \\
  cT3 &  & 0.347 \\
  &  & (0.236) \\
  cT1 &  & $-$0.248 \\
  &  & (0.271) \\
  cN2 & 0.032 & 0.387 \\
  & (0.335) & (0.369) \\
  cN3 &  & $-$0.252 \\
  &  & (0.730) \\
  cN0 & $-$0.009 & $-$1.237$^{***}$ \\
  & (0.322) & (0.260) \\
  age & $-$0.017 & 0.012 \\
  & (0.011) & (0.010) \\
  sex0 & $-$0.225 & 0.401$^{*}$ \\
  & (0.345) & (0.238) \\
  G0,1,2 & $-$0.141 &  \\
  & (0.231) &  \\
  G2 &  & $-$0.600$^{***}$ \\
  &  & (0.226) \\
  G1 &  & $-$1.167$^{*}$ \\
  &  & (0.652) \\
  ASA1 & 0.104 & 0.137 \\
  & (0.254) & (0.315) \\
  ASA3\&4 & 0.533$^{*}$ & $-$0.339 \\
  & (0.316) & (0.324) \\
 \hline \\[-1.8ex]
Observations & 150 & 162 \\
R$^{2}$ & 0.481 & 0.593 \\
Max. Possible R$^{2}$ & 0.996 & 0.998 \\
Log Likelihood & $-$361.742 & $-$450.474 \\
Wald Test & 98.000$^{***}$ (df = 9) & 131.470$^{***}$ (df = 12) \\
LR Test & 98.460$^{***}$ (df = 9) & 145.467$^{***}$ (df = 12) \\
Score (Logrank) Test & 109.909$^{***}$ (df = 9) & 164.954$^{***}$ (df = 12) \\
\hline
\hline \\[-1.8ex]
\textit{Notes:}  & \multicolumn{2}{r}{$^{*}$p$<$0.1; $^{**}$p$<$0.05; $^{***}$p$<$0.01} SE in parentheses \\
\end{tabular}
\end{table*}

Esophageal Cancer (EC) is the eighth most common cancer worldwide, with less than 20\% of patients surviving more than five years. Whereas in western countries the number of esophageal squamous cell carcinoma (ESCC) is declining, the number of adenocarcinomas of the esophagogastric junction (AEG) diagnosed is increasing \citep{bray2018global}.

Despite the development of multimodal approaches, combining surgical resection with perioperative chemo-(radio)therapy, 5-year survival rate for patients diagnosed with AEG remains poor \citep{behrens2015perioperative}. Inflammation, as one of the hallmarks of cancer, is an acknowledged factor in tumor biology. Inflammation-driven tumorigenesis and tumor progression plays a crucial role in malignant diseases. Systemic inflammatory response (SIR) in the context of tumor-associated inflammation has been demonstrated to diminish outcome and be of major prognostic importance in various cancers. Investigation of tumor-driving inflammation-based components is of major importance and targeting pathways of inflammatory response might become a cornerstone of cancer treatment \citep{hanahan2011hallmarks}.

Therefore, identification of easy-available markers might help to determine individual treatment approaches. The utility of inflammation-based scores, such as the systemic immune-inflammation index (SII), are based
on routinely obtained markers that are available before surgery. Elevated SII has been reported to be associated with clinico-pathological parameters and has been proven to be an independent prognostic factor in a number of malignancies \citep{aziz2019systemic, hong2015systemic, hu2014systemic}.

In the present application, we investigate the prognostic value of SII in patients with adenocarcinoma of the gastroesophageal junction who underwent primary resection or were treated with neoadjuvant therapy before surgery. The data set consists of the survival time (in months) of 287 consecutive patients that underwent curative resection of locally advanced adenocarcinomas of the gastroesophageal junction between January 1992 and April 2016 at the Department of Surgery at the Medical University of Vienna were identified from a prospectively maintained database.
Serum concentrations of platelets, neutrophils, and lymphocytes were measured within 3 days before the start of neoadjuvant treatment or surgery in patients treated with primary surgery. The SII is calculated as follows: $\text{SII} = \frac{\text{platelet}*\text{neutrophil}}{\text{lymphocyte}}$. In addition to the SII several common control variables described in Table \ref{tab:variables} are used for estimation. The sampler is run for $100.000$ iterations with a burn-in of $20.000$ iterations and a thinning factor of 20. Both $\sqrt \theta_k$ and $\beta_k$ are placed under a triple gamma prior, with the default hyperparameters from \cite{winkler2022shrinkdsm}. The partitioning set $S$ contains every observed death.

The plots in the first row of Figure \ref{fig:posteriors} show the posterior parameter associated with SII for patients who were treated with primary surgery and neoadjuvant treatment, respectively. The shaded blue area represent the pointwise 50\% and 95\% credible intervals. The risk associated with SII increases over the first 10 months for both groups after starting treatment (either surgery for primary resected patients or neoadjuvant treatment) and decreases after the initial peak. However, while the parameter becomes statistically indistinguishable from $0$ for patients who underwent primary resection after $\sim 6$ years, it stays constant with a median around $1$ for those who received neoadjuvant therapy. A Cox proportional-hazards model estimated with the same data set (Table \ref{tab:coxph}) is unable to capture this time-variation and simply results in a smaller estimate for primary resected patients. Thus the proposed model results in two important interpretative differences: First, the parameters associated with SII for the two groups are practically indistinguishable for the first two years. Second, while SII remains an important factor for survival in patients receiving neoadjuvant chemotherapy, it is only important for the first $6$ years or so for those undergoing primary resection.

To showcase that constant parameters can also be captured with the proposed model, the posterior parameters associated with the clinical lymph node stage (cN) are also included in Figure \ref{fig:posteriors}. All parameters that appear as statistically insignificant at the standard 5\% level according to the Cox PH model are also shrunk to $0$ throughout the entire observed time period using our Bayesian approach. Furthermore, the parameter associated with cN = $0$ is significantly below $0$ in the Cox PH model ($z-score=-4.761$). In the proposed model the vast majority of posterior mass is around $-1$ and displays virtually no time-variation.

\begin{figure}
\centering
\includegraphics[width=0.66\maxwidth]{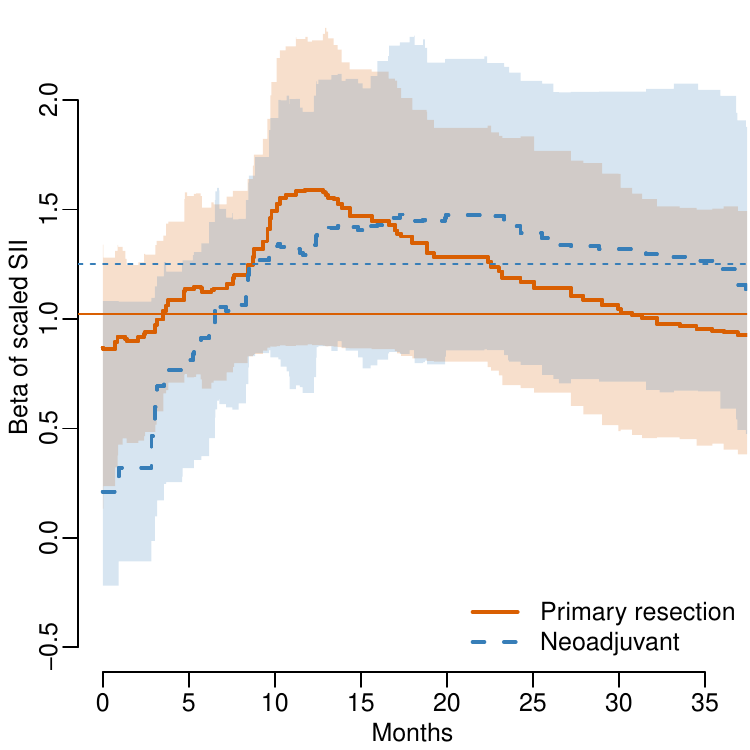}
\caption{Posterior distribution of the coefficient associated with SII parameter for primary resected and neoadjuvant treated patients in the first 36 months. Lines represent the posterior median, while shaded regions represent 95\% credible intervals. Horizontal lines represent the corresponding estimates of the Cox proportional-hazards model.}
\label{fig:overlap}
\end{figure}

Crucially, the time-variation afforded by the model also allows for more nuanced interpretations of the effect of SII on survival time. Figure~\ref{fig:overlap} shows the posterior distribution of the coefficient associated with SII for primary resected and neoadjuvant treated patients in the first 36 months. Were a practitioner to guide their attention to patients based solely on the estimates from the Cox proportional hazards model, they would conclude that those that undergo primary resection are at higher risk than those that receive neoadjuvant treatment. However, in approximately the first 10 months, the model gives credence to the opposite conclusion, with the effect reversing around the 17 month mark. Especially at the beginning of treatment, this may have important implications for outcomes of patients.  

This example highlights the importance of the time-varying parameter approach when it comes to understanding the effect of SII on survival time. The proposed model is shown to be flexible enough to capture both constant and time-varying parameters while also giving a clear indication of which parameters should be included in the model without the need for much hyperparameter tuning.

\section{Conclusion} \label{sec:conc}

This paper presents a flexible yet sparse Bayesian model for survival analysis, combining a piecewise-exponential hazard model with a shared frailty term and modern shrinkage priors. This provides theoretical and practical advantages.
The flexible model structure allows for time-variation in the effects of covariates, which is crucial in many real-world applications in medicine.
In existing methods, this flexibility often comes at the cost of requiring extensive hyper-parameter tuning or prior knowledge about the parameter space to make estimation possible.
However, the triple gamma shrinkage prior applied in our model can learn the level of sparsity directly from the data while still allowing researchers to easily control shrinkage by setting interpretable hyperparameters if so desired. 
Our model applies shrinkage to both the covariate effects on the hazard and the variance of the innovations governing time variation, thereby identifying which covariates affect survival and determining whether their impacts evolve over time.
The presented simulation study shows that it outperforms existing methods for time-varying effects across different levels of sparsity, dimensionality, and sample size, thereby highlighting the practical advantages of the proposed model. 
Furthermore, even in simulations with constant parameters, it remains among the best-performing methods. This demonstrates that our approach can effectively balance flexibility and model simplicity, allowing researchers to   apply the model across various settings while avoiding restrictive assumptions about the underlying data-generating process (e.g., constant parameters).
The application to a data set of patients with adenocarcinoma of the gastroesophageal junction demonstrates the practical advantages of the model's flexibility. Comparing the effects of SII on neoadjuvant-treated and primary resected patients, the time variation in effects reveals important heterogeneity not captured by models assuming constant hazard.
The proposed model is implemented in the R package \texttt{shrinkDSM}, available on CRAN, providing an easy-to-use tool for researchers. The efficient MCMC algorithm is fully implemented in \texttt{C++}, ensuring reasonably fast sampling.

In summary, we present an efficient model for survival analysis that combines ease of use with flexibility and strong theoretical underpinnings. The model is particularly well-suited for applications in which researchers are not able to make strong assumptions about the underlying data-generating process.

\bibliographystyle{apalike}

\bibliography{./mybib_fixed}



\newpage

\appendix


\noindent
\textbf{\LARGE Appendix}

\section{MCMC Algorithm}

Draws from the posterior distribution are obtained through an MCMC Gibbs sampler with MH steps. The MCMC scheme extends those developed in \cite{wagner2011bayesian}, \cite{bitto2019achieving}, \cite{cadonna2020triple} and \cite{knaus2021shrinkage}. The core idea behind the sampler is to use the data augmentation scheme described in Section~\ref{sec:dataug} to transform the model into a conditionally Gaussian state space model and then use the many available tools for this class of models to sample from the posterior distribution. 

For sampling from the conditionally Gaussian state space model, three things are important to note: First, to improve mixing, the sampler uses both the parameterization in \eqref{eq:gssm}, as well as the non-centered parameterization of \cite{fruhwirth2010stochastic}:
\begin{equation} \label{eq:noncen}
  \begin{aligned}
  \bm{x}_{j}&=\bm{Z}_{j}\bm{\beta} + \bm{Z}_{j} \bm{Q}(\bm{\theta})^{1/2} \tilde{\bm{\beta}}_{j}+\bm{\varepsilon}_{j}, \quad \bm{\varepsilon}_{j} \sim N\left(\bm{0}, \bm{V}_{j}\right), \\
  \tilde{\bm{\beta}}_{j}&=\tilde{\bm{\beta}}_{j-1}+\bm{w}_{j}, \quad \bm{w}_{j} \sim N(\bm{0}, \bm{I}),
  \end{aligned}
\end{equation}

where $\bm{Q}(\bm{\theta})^{1/2} = \text{diag}(\sqrt{\theta}_1, \dots, \sqrt{\theta}_K)$. $\bm \beta_j$ and $\tilde{\bm \beta}_j$ are linked through the simple transformation
$$
\bm \beta_j = \bm \beta + \bm Q(\bm \theta)^{1/2} \tilde{\bm \beta}_j.
$$
This transformation implies the following prior on $\sqrt{\theta}_k$:
\begin{equation*}
  \begin{gathered}
    \sqrt \theta_k | \tau^\theta, \lambda_k^\theta  \sim \mathcal N \left(0, \lambda_k^\theta \tau^\theta\right), \quad \lambda_k^\theta \sim F(2a^\theta, 2c^\theta) \\
    \tau^\theta \sim F(2c^\theta, 2a^\theta).
  \end{gathered}
\end{equation*}
Second, the sampler uses different representations of the triple gamma, normal-gamma-gamma and F prior, as this facilitates Gibbs computations \citep{cadonna2020triple}. The representations for the triple gamma and normal-gamma-gamma prior are as follows:
\begin{equation} \label{eq:triple_expanded}
  \begin{gathered}
  \beta_{k} | \xi_k^\beta, \kappa_k^\beta, \tilde\tau^\beta, a^\beta, c^\beta \sim \mathcal N \left(0, \frac{2 c^\beta \xi_k^\beta}{\tilde\tau^\beta a^\beta \kappa_k^\beta}\right) \\
  \xi_k^\beta | a^\beta \sim \mathcal G \left(a^\beta, 1\right), \quad \kappa_k^\beta | c^\beta \sim \mathcal G \left(c^\beta, 1\right), \\
  \theta_k | \xi_k^\theta, \kappa_k^\theta, \tilde\tau^\theta, a^\theta, c^\theta  \sim \mathcal G \left(\frac 1 2, \frac{\tilde\tau^\theta a^\theta  \kappa_k^\theta}{4 c^\theta \xi _k^\theta}{}\right) \\
  \xi_k^\theta | a^\theta \sim \mathcal G \left(a^\theta, 1\right) \quad \kappa_k^\theta | c^\theta \sim \mathcal G \left(c^\theta, 1\right).
\end{gathered}
\end{equation}
For notational simplicity, we define 
$$\phi_k^\beta = \frac{2 c^\beta \kappa_k^\beta}{\tilde\tau^\beta a^\beta \xi_k^\beta}, \quad \phi_k^\theta = \frac{2 c^\theta \kappa_k^\theta}{\tilde\tau^\theta a^\theta \xi_k^\theta}.
$$
The expanded representation of the F prior, for both $\tilde\tau^\beta$ and $\tilde\tau^\theta$, is given by
\begin{equation} \label{eq:f_expanded}
  \begin{gathered}
    \tilde\tau | a, d_2 \sim \mathcal G \left(a, d_2\right), \quad d_2 | a, c \sim \mathcal G \left(c, \frac{2 c} {a}\right).
  \end{gathered}
\end{equation}
Note that, in this representation, $\tilde\tau$ marginally follows a scaled version of the prior on $\tau$:
$$
\frac 2 {\tilde \tau} \sim F(2c, 2a).
$$
Despite this, the representation of the prior structure in \eqref{eq:triple_expanded} and \eqref{eq:f_expanded} is equivalent to the representation in \eqref{eq:tg_prior} and \eqref{eq:ngg_prior}, however  it allows for easier derivation of closed-form expressions for the full conditionals. 

Third, to improve mixing for the sampling of the parameters $a^\beta, a^\theta, c^\beta$ and $c^\theta$, we use partially marginalized representations of the triple gamma prior for the Metropolis-within-Gibbs steps \citep{cadonna2020triple}. When marginalizing over $\kappa_k^\beta$ and $\kappa_k^\theta$, this results in
\begin{equation} \label{eq:tg_marg}
  \begin{gathered}
\beta_k|\xi_k^\beta, a^\beta, c^\beta, \tilde\tau^\beta \sim t_{2c^\beta}\left(0, \frac{2\xi_k^\beta}{a^\beta\tilde\tau^\beta}\right), \\ \xi_k^\beta | a^\beta \sim \mathcal G \left(a^\beta, 1\right), \\
\sqrt \theta_k|\xi_k^\theta, a^\theta, c^\theta, \tilde\tau^\theta \sim t_{2c^\theta}\left(0, \frac{2\xi_k^\theta}{a^\theta\tilde\tau^\theta}\right), \\ \xi_k^\theta | a^\theta \sim \mathcal G \left(a^\theta, 1\right).
\end{gathered}
\end{equation}
When marginalizing over $\xi_k^\beta$ and $\xi_k^\theta$ the density of $\beta_k$ and $\sqrt \theta_k$ is available in closed, albeit not well-known form. More specifically:
\begin{align} \label{eq:ngg_marg}
p(\beta_k&|\kappa_k^\beta, a^\beta, c^\beta, \tilde\tau^\beta) = \frac{\sqrt{\frac{\tilde\tau^\beta\kappa_k^\beta a^\beta}{c^\beta}}^{a^\beta + 1/2}}{\Gamma(a^\beta)\sqrt\pi 2^{a^\beta - 1/2}} |\beta_k|^{a^\beta - 1/2} \\
&\times K_{a^\beta - 1/2}\left(\sqrt{\frac{\tilde\tau^\beta\kappa_k^\beta a^\beta}{c^\beta}}|\beta_k|\right),
\end{align}
where $K_\nu(z)$ is the modified Bessel function of the second kind. The density of $\sqrt \theta_k$ is analogous.
Given these details, it is now possible to summarize the MCMC algorithm in Algorithm~\ref{algo_basic}.

\begin{algorithm}
  \small  
  \begin{enumerate}

    \item[\textit{a)}] Sample the latent states $\tilde{\bm \beta}_0, \dots, \tilde{\bm \beta}_J$ in the non-centered parameterization using the precision-based method of \cite{mccausland2011simulation}.
    \item[\textit{b)}] Conditional on $\tilde{\bm{\beta}}, \bm V_1, \dots, \bm V_J, \bm \phi^\beta, \bm \phi^\theta$ and $\bm x_1, \dots, \bm x_J$, the non-centered parameterization in \eqref{eq:noncen} can be written as a linear regression model:
    $$
    \bm x = \bm M \bm \alpha + \bm \varepsilon, \quad \varepsilon \sim \mathcal N \left(\bm 0, \bm V \right),
    $$
    where 
    $$
    \bm x = 
    \begin{pmatrix}
      \bm x_1 \\
      \vdots \\
      \bm{x}_{J}
      \end{pmatrix},
      \
    \bm M = 
    \begin{pmatrix}
      z_{11}, \dots, z_{K1} & z_{11} \tilde \beta_{11}, \dots \tilde \beta_{K1} \\
      \vdots  & \vdots \\
      z_{1{n_J}}, \dots, z_{K{n_J}} & z_{1{n_J}} \tilde \beta_{1J}, \dots \tilde \beta_{KJ} 
    \end{pmatrix},
    \
    \bm \varepsilon = 
    \begin{pmatrix}
      \bm \varepsilon_1, \\
      \vdots \\
      \bm \varepsilon_J 
    \end{pmatrix},
    \ 
    \bm V = 
    \begin{pmatrix}
      \text{diag}(\bm V_1), \\
      \vdots \\
      \text{diag}(\bm V_J) 
    \end{pmatrix},
    $$
    and, crucually, $\bm \alpha = (\beta_1, \dots, \beta_K, \sqrt \theta_1, \dots, \theta_K)'$. In this conditional representation, the prior on $\bm \alpha$ is 
    $$
    \bm \alpha | \bm \phi^\beta, \bm \phi^\theta \sim \mathcal N \left(\bm 0, \bm G\right), \quad \bm G = \text{diag}\left(\phi_1^\beta, \dots, \phi_K^\beta, \phi_1^\theta, \dots, \phi_K^\theta\right).
    $$
    This implies the following conditional multivariate Gaussian posterior for $\bm \alpha$ to sample from: 
    $$
    \bm \alpha | \tilde{\bm{\beta}}, \bm V_1, \dots, \bm V_J, \bm \phi^\beta, \bm \phi^\theta, \bm x \sim \mathcal N \left(\bm a, \bm A\right),
    $$
    where
    $$
    \bm a = \bm A \bm M' \bm V^{-1} \bm x, \quad \bm A = (\bm M' \bm V^{-1} \bm M + \bm G^{-1})^{-1}.
    $$
    \item[\textit{c)}] Re-sample $\beta_1, \dots, \beta_K$ and $\theta_1, \dots, \theta_K$ from the centered parameterization through the following steps:
    \begin{enumerate}
      \item[\textit{c.1)}] Transform the draws of the non-centered states $\tilde{\bm{\beta}}_{0}, \dots, \tilde{\bm{\beta}_J}$ back into their centered counterparts via the transformation
      $$
      \beta_{kj} = \beta_k + \sqrt \theta_k \tilde{\beta}_{kj}.
      $$
      Further, store the signs of $\sqrt \theta_1, \dots, \sqrt \theta_K$. 
      \item[\textit{c.2)}] For $k = 1, \dots, K$, sample $\theta_k$ from the generalized inverse Gaussian posterior
      $$
      \theta_k | \beta_{k0}, \dots, \beta_{kJ}, \beta_k, \phi_k^\theta | \sim \mathcal{GIG} \left(-\frac J 2, \frac 1 {\phi_k^\theta}, \sum_{j=1}^J (\beta_{kj} - \beta_{k,j-1})^2 + (\beta_{k0} - \beta_k)^2\right).
      $$
      \item[\textit{c.3)}] For $k = 1, \dots, K$, sample $\beta_k$ from the Gaussian posterior
      $$
      \beta_k|\beta_{k0}, \theta_k, \phi_k^\beta \sim \mathcal N\left(\frac{\beta_{k0}\phi_k^\beta}{\phi_k^\beta + \theta_k}, \frac{\phi_k^\beta\theta_k}{\phi_k^\beta + \theta_k}\right).
      $$
      \item[\textit{c.4)}] For $k = 1, \dots, K$, determine $\sqrt \theta_k^{\ new}$ by taking the square root of $\theta_k$ and assigning the stored sign from step \textit{d.1)}. Generate the updated draws of $\tilde{\bm \beta}_{0}, \dots, \bm \tilde{\bm \beta}_{J}$ throught the transformation
      $$
      \tilde \beta_{kj}^\text{new} = \left(\beta_{kj} - \beta_k\right)/\sqrt \theta_k^{\ new}.
      $$
    \end{enumerate}
  \end{enumerate}
  \caption{MCMC inference for normal dynamic survival models under triple gamma shrinkage prior.}
  \label{algo_basic}
  \end{algorithm}


  \begin{algorithm}
    \small
    \begin{enumerate}
      \item[\textit{d)}] Sample $a^\beta|\beta_1, \dots, \beta_K, \kappa_1, \dots, \kappa_K^\beta, c^\beta, \tilde\tau^\beta, $ and $a^\theta|\theta_1, \dots, \theta_K, \kappa_1, \dots, \kappa_K^\theta, c^\theta, \tilde\tau^\theta$ through a random walk Metropolis step, using the marginalized representation in \eqref{eq:ngg_marg}. 
      \item[\textit{e)}] For $k = 1, \dots, K$, sample $\xi_k^\beta$ from a generalized inverse Gaussian distribution
    $$
    \xi_k^\beta | \beta_k, \kappa_k^\beta, a^\beta, c^\beta,  \tilde\tau^\beta \sim \mathcal{GIG}\left(a^\beta - \frac 1 2, 2, \frac{\beta_k^2 \kappa_k^\beta\tilde\tau^\beta a^\beta}{2 c^\beta}\right).
    $$
      \item[\textit{f)}] For $k = 1, \dots, K$, sample $\xi_k^\theta$ from a generalized inverse Gaussian distribution
      $$
      \xi_k^\theta | \theta_k,\kappa_k^\theta, a^\theta, c^\theta,  \tilde\tau^\theta \sim \mathcal{GIG}\left(a^\theta - \frac 1 2, 2, \frac{\theta_k \kappa_k^\theta\tilde\tau^\theta a^\theta}{2 c^\theta}\right).
      $$
      \item[\textit{g)}] Sample $c^\beta|\beta_1, \dots, \beta_K, \xi_1, \dots, \xi_K^\beta, a^\beta, \tilde\tau^\beta$ and $c^\theta|\theta_1, \dots, \theta_K, \xi_1, \dots, \xi_K^\theta, a^\theta, \tilde\tau^\theta$ with a random walk Metropolis step, using the marginalized student t representation in \eqref{eq:tg_marg}. 
      \item[\textit{h)}] For $k = 1, \dots, K$, sample $\kappa_k^\beta$ from a gamma distribution
      $$
      \kappa_k^\beta | \beta_k, \xi_k^\beta, a^\beta, c^\beta, \tilde\tau^\beta \sim \mathcal G\left(\frac 1 2 + c^\beta, \frac{\beta_k^2\tilde\tau^\beta a^\beta}{4 c^\beta\xi_k^\beta} + 1\right).
      $$
    \item[\textit{i)}] For $k = 1, \dots, K$, sample $\kappa_k^\theta$ from a gamma distribution
  $$
  \kappa_k^\theta | \theta_k, \xi_k^\theta, a^\theta, c^\theta, \tilde\tau^\theta \sim \mathcal G\left(\frac 1 2 + c^\theta, \frac{\theta_k\tilde\tau^\theta a^\theta}{4 c^\theta\xi_k^\theta} + 1\right).
  $$
  \item[\textit{j)}] Sample $d_2^\beta$ from $d_2^\beta | a^\beta, c^\beta, \tilde\tau^\beta \sim \mathcal G\left(a^\beta + c^\beta, \tilde\tau^\beta + \frac{2c^\beta}{a^\beta}\right)$ and then sample $\tilde\tau^\beta$ from a gamma distribution
  $$
  \tilde\tau^\beta | \{\beta_k, \xi_k^\beta, \kappa_k^\beta\}_{k=1}^K, a^\beta, c^\beta, d_2^\beta \sim \mathcal G \left(\frac{d_2^\beta}{2} + a^\beta, \frac{a^\beta}{4c^\beta}\sum_{k=1}^K\frac{\kappa_k^\beta}{\xi_k^\beta}\beta_j^2 + d_2^\beta\right).
  $$
  \item[\textit{k)}] Sample $d_2^\theta$ from $d_2^\theta | a^\theta, c^\theta, \tilde\tau^\theta \sim \mathcal G\left(a^\theta + c^\theta, \tilde\tau^\theta + \frac{2c^\theta}{a^\theta}\right)$ and then sample $\tilde\tau^\theta$ from a gamma distribution
  $$
  \tilde\tau^\theta | \{\theta_k, \xi_k^\theta, \kappa_k^\theta\}_{k=1}^K, a^\theta, c^\theta, d_2^\theta \sim \mathcal G \left(\frac{d_2^\theta}{2} + a^\theta, \frac{a^\theta}{4c^\theta}\sum_{k=1}^K\frac{\kappa_k^\theta}{\xi_k^\theta}\theta_j + d_2^\theta\right).
  $$
  \item[\textit{l)}] Sample the auxiliary variables $\tau_{ij}$ and the component indicators $r_{ij}$ as in \cite{wagner2011bayesian}, Section 3.4, step (c) and use these to update $\bm x_j$, for $j = 1, \dots, J$.
  \end{enumerate}
  \caption*{(cont.) MCMC inference for normal dynamic survival models with shrinkage and a factor.}
  \end{algorithm}
 
  \newpage

\section{Additional MCMC steps for the factor component}

\begin{algorithm}[]
  \small
       \begin{enumerate}
       \item[\textit{a)}] Use steps \textit{a - k} from Algorithm~\ref{algo_basic}.
       \item[\textit{b)}] Define the following stacked counterparts for parameters used in the model:
       \begin{itemize}
            \item $\tilde {\boldsymbol x}_j = {\boldsymbol x}_j - {\boldsymbol z}_j{\boldsymbol \beta}_j$, $\tilde {\boldsymbol{x}} = [\tilde {\boldsymbol{x}}_1, \dots \tilde {\boldsymbol{x}}_J]^\prime$,
            \item ${\boldsymbol{V}} = [{\boldsymbol{V}}_1, \dots {\boldsymbol{V}}_J]^\prime$,
            \item $\boldsymbol f = [\underbrace{(f_1, \dots, f_1)}_{n_1 \text{ elements}}, \dots, \underbrace{(f_J, \dots, f_J)}_{n_J \text{ elements}}]^\prime$,
       \end{itemize}
       and the index $p$, for $p = {1, \dots \sum_{j=1}^J n_j}$ with group specific indices $p_g$, with $p_g = \{p | \text{ individual }i \text{ in group } g \}$. These effectively select the rows belonging to one group from the stacked counterparts of the parameters. Then sample $\phi_g$ for $g = 1,\dots,G$ from the following conditional posterior:
       $$
       \begin{aligned}
            p(\phi_g|\boldsymbol y, \boldsymbol z, \boldsymbol \theta_{-\phi_g}) & \propto p(\tilde {\boldsymbol x}_g|\phi_g, {\boldsymbol{V}, \boldsymbol f}) p(\phi_g) \\
            &\propto exp\left(\frac{1}{2\bar\sigma^2_{\phi_g}} \left(\phi_g - \bar\sigma^2_{\phi_g}\sum_{e:e\in p_g}\tilde {\boldsymbol x}_e\boldsymbol f_e/\boldsymbol V_e \right)^2 \right)
            \end{aligned}
       $$
     with $\bar\sigma^2_{\phi_g} = \left(\frac{1}{\sum_{e: e \in p_g} \boldsymbol f_e^2/\boldsymbol V_e} + \frac{1}{\sigma^2_{\phi_g}}\right)$.
  
  \item[\textit{c)}] Sample $f_j$ for $j = 1, \dots, J$ by defining the following $J$ univariate regression problems:
      $$
      \tilde {\boldsymbol x}_j = \boldsymbol\phi_j f_j + \boldsymbol\varepsilon_j,
      $$
      where $\boldsymbol\phi_j$ is a vector of the stacked values of $\phi_g$ for all individuals $i$ still alive at time $j$. Posterior samples are generated by drawing from the resulting conditional normal posteriors.
     \item[\textit{d)}] Perform a boosting step to improve the mixing of the factor loadings.
     \begin{enumerate}
       \item Define $\phi_m$ as the largest of $\phi_1, \dots \phi_G$ in absolute value, $\phi_g^* = \frac{\phi_g}{\phi_m}$ and $f_j^* = f_j \phi_m$.
       \item Sample $\phi_m^2$ from the full conditional posterior in Section~\ref{full_cond_phi_m}. Define $\phi_m^\text{new}$ as $\phi_m^\text{new} = \sqrt{\phi_m^2}$, preserving the sign of the original $\phi_m$.
       \item Determine $\phi_g^{\text{new}}$ through the transformation $\phi_g^{\text{new}} = \phi_g^*\phi_m^\text{new}$, for $g = 1, \dots, G$. Furthermore, determine $f_j^{\text{new}}$ through the transformation $f_j^{\text{new}} = f_j^* / \phi_m^\text{new}$, for $j = 1, \dots, J$.
     \end{enumerate}
     \item[\textit{e)}] Sample the variances of the factor loadings $\tau^\phi_g$ for $g = 1, \dots, G$ as in step $4$ from Algorithm 1 in \cite{knaus2021shrinkage}.
     \item[\textit{f)}] Sample $f_j$ for $j = 1, \dots, J$ by defining the following $J$ univariate regression problems:
      $$
      \tilde {\boldsymbol x}_j = \boldsymbol\phi_j f_j + \boldsymbol\varepsilon_j,
      $$
      where $\boldsymbol\phi_j$ is a vector of the stacked values of $\phi_g$ for all individuals $i$ still alive at time $j$. Posterior samples are generated by drawing from the resulting conditional normal posteriors.
      \item[\textit{g)}] Sample the persistence $\phi_f$, the volatility of the volatility $\sigma^2_f$ and the log-volatilities $h =(h_0,\dots,h_J)$ as in \cite{kastner2016dealing} using \texttt{stochvol} \cite{hosszejni2021modeling}.
       \item[\textit{h)}] Sample the auxiliary variables $\tau_{ij}$ and the component indicators $r_{ij}$ as in \cite{wagner2011bayesian}, Section 3.4, step (c) and use these to update $\boldsymbol x_j$, for $j = 1, \dots, J$.
  \end{enumerate}
  
        \caption{MCMC inference for normal dynamic survival models with shrinkage and a factor.}
        \label{algo_factor}
  
  \end{algorithm}

\subsection{Full conditional posterior of $\phi_m$}
\label{full_cond_phi_m}

Let $\phi_m$ be the largest of $\phi_1, \dots \phi_G$ in absolute value and define the following quantities:
$$
\phi_g^* = \frac{\phi_g}{\phi_m}, \quad f_j^* = f_j \phi_m.
$$

This implies the following working priors:
$$
\phi^2_m \sim \mathcal G \left(\frac{1}{2}, \frac{1}{2\sigma^2_{\phi_m}}\right), \quad f_j^* \sim \mathcal N(0, \exp(h_j)\phi_m^2), \quad \phi_g^* \sim \mathcal N \left(0, \frac{\sigma^2_{\phi_g}}{\phi_m^2}\right),
$$
which, in turn, leads to the following full conditional posterior for $\phi_m^2$:
\begin{equation}
\begin{aligned}
p(\phi_m^2|\phi_g^*, h_1, \dots h_J, &f_1^*, \dots f_J^*) \propto \prod_{g:g\neq m} p(\phi_g^*|\phi_m^2)\prod_{j=1}^J p(f_j^*|\phi_m^2) p(\phi_m^2) \\
& \propto \left[\prod_{g:g\neq m} (\phi_m^2)^\frac{1}{2} exp\left(-\frac{1}{2}\frac{\phi_g^{*2}}{\sigma^2_{\phi_g}}\phi_m^2\right)\right] \left[\prod_{j=1}^J (\phi_m^2)^{-\frac{1}{2}}exp\left(-\frac{1}{2}\frac{f_j^{*2}}{\exp(h_j)}\frac{1}{\phi_m^{2}}\right)\right] \\
&\times (\phi_m^2)^{\frac{1}{2}-1}exp\left(-\phi_m^2\frac{1}{2\sigma^2_{\phi_m}}\right) \\
&=(\phi_m^2)^{\frac{G-J}{2}- 1} exp\left[-\frac{1}{2}\left(\phi_m^2\left(\frac{1}{\sigma^2_{\phi_m}} + \sum_{g:g\neq m} \frac{\phi_g^{*2}}{\sigma^2_{\phi_g}} \right) + \frac{1}{\phi_m^2}\sum_{j=1}^J\frac{f^{*2}_j}{\exp(h_j)}\right)\right],
\end{aligned}
\end{equation}
which is the kernel of a $GIG\left(\frac{G-J}{2}, \frac{1}{\sigma^2_m} + \sum_{g:g\neq m} \frac{\phi_g^{*2}}{\sigma^2_g}, \sum_{j=1}^J\frac{f^{*2}_j}{\exp(h_j)}\right)$ distribution.

\end{document}